\documentstyle[amssymb,preprint,eqsecnum,aps,prb,12pt,amstex]{revtex}

\begin{document}
\title{Instability of metal-insulator transition against thermal cycling in phase
separated Cr-doped manganites }
\author{R. Mahendiran$^{1}$, B. Raveau, M. Hervieu, C. Michel, and A. Maignan}
\address{Laboratoire CRISMAT, ISMRA, Universit\'{e} de Caen, 6 Boulevard du\\
Mar\'{e}chal Juin, 14050 Caen-Cedex, France.}
\maketitle

\begin{abstract}
Pr$_{0.5}$Ca$_{0.5}$Mn$_{1-x}$Cr$_{x}$O$_{3}$ ($x$= 0.015-0.05) undergoes an
insulator-metal (I-M) transition below a temperature T$_{p}$ driven by
percolation of ferromagnetic metallic clusters in a charge ordered
insulating matrix. Surprisingly, the I-M transition in these compounds is
unstable against thermal cycling : T$_{p}$ decreases and the resistivity at T%
$_{p}$ increases upon temperature cycling between a starting temperature T$%
_{S}$ and a final temperature T$_{F}$ and changes are larger for smaller x.
The resistivity transition to the low temperature metallic state in x =
0.015 can be completely destroyed by thermal cycling in absence of magnetic
field as well as under H = 2 T. Magnetization measurements suggest that the
ferromagnetic phase fraction decreases with increasing thermal cycling. We
suggest that \ the increase of strains in the ferromagnetic-charge ordered
interface could be a possible origin of the observed effect.

PACS NO: 75.30.Vn, 71.30+h, 75.30. +Kz, 77.50. -y.
\end{abstract}

\newpage

\section{INTRODUCTION}

Over the past few years, manganites of the general formula RE$_{1-x}$AE$_{x}$%
MnO$_{3}$ have been studied extensively\cite{Ramirez}. Although primary
interest in these compounds is motivated by the discovery of colossal
magnetoresistance, understanding the physical properties even in absence of
magnetic field is still far from clear. In particular, the behavior of \ the
resistivity ($\rho $) is intriguing. The well studied La$_{0.67}$Ca$_{0.33}$%
MnO$_{3}$ compound with hole density of n$_{h}$ = 0.33 per Mn exhibits a
paramagnetic insulator to a ferromagnetic metal transition upon cooling from
high temperature at T$_{p}$ = T$_{C}$ where T$_{C}$ is the ferromagnetic
Curie temperature. For the same hole concentration, the residual resistivity
spans over six orders of magntitude and T$_{p}$ shifts down, either with
different choices of the rare earth (RE) and the alkaline (AE) ions\cite
{Coey} or by only varying their fraction as in (La$_{1-y}$Pr$_{y}$)$_{0.67}$%
Ca$_{0.33}$MnO$_{3}$\cite{Hwang}. The paramagnetic state of some other
manganites like La$_{0.7}$Sr$_{0.3}$MnO$_{3}$ is metallic\cite{Urushibara}
instead of insulating. Metallic like resistivity behavior (d$\rho $/dT 
\mbox{$>$}%
0) in the paramagnetic state was also reported in RE$_{1-x}$Ca$_{x}$MnO$_{3}$
for x = 0.8-0.98 and RE = Pr, Nd, Sm, etc \cite{Maignan}. Charge
localization due to random Coulomb potential of RE$^{3+}$, AE$^{2+}$ ions
and spin disorder scattering alone are inadequate to understand these
results and other polaronic mechanisms involving strong electron-phonon-spin
coupling have to be considered\cite{Millis1}.

\bigskip

One of the peculiar aspects of the manganites with high residual resistivity
is that T$_{p}$ occurs much below T$_{C}$ in contradiction to the double
exchange mechanism which associates the onset of metallic conduction to the
ferromagnetic alignment of Mn spins (T$_{p}$ = T$_{C}$). There are growing
experimental evidences that in most, if not all of the compounds with lower T%
$_{P}$ (%
\mbox{$<$}%
150 K), insulator-metal transition is driven by percolation of \ the
ferromagnetic metallic clusters in either paramagnetic or charge ordered
insulating background\cite{Ibarra, Moreo,Mahi1,Moritomo,Kimura}.The huge
hysteresis in resistivity of these percolation driven compounds suggests
that the transition is of first order. However, kinematics of the transition
is largely unknown except for two cases. Babushkina et al\cite{Babushkina}
found that low the temperature zero field resistivity of (La$_{0.5}$Nd$%
_{0.5} $)$_{0.7}$Ca$_{0.3}$MnO$_{3}$ thin film relaxes to lower values as a
function of time. Anane et al\cite{Anane} reported that the magnetic field
driven metallic state of Pr$_{0.67}$Ca$_{0.33}$MnO$_{3}$ is unstable and the
resistivity at a given temperature jumps abruptly from a low to a high value
at a particular value of time after the removal of \ external magnetic
field. The transition in both of these compounds is of isothermal type,
involving growth of ferromagnetic metallic nuclei as a function of time in a
charge ordered matrix in the former compound and vice versa in the latter.
It is not clear whether time dependent growth of \ the low temperature phase
is the general aspect of the first order transition in manganites. In this
back ground, we have chosen Pr$_{0.5}$Ca$_{0.5}$Mn$_{1-x}$Cr$_{x}$O$_{3}$
compounds which are best known for the spectacular insulator-metal
transition shown by them in absence of a magnetic field\cite{Raveau}. The
insulator -metal transition in these compounds is due to the percolation of
the ferromagnetic clusters in the charge ordered antiferromagnetic matrix
and the ferromagnetic phase fraction can be controlled by changing the
composition (x) as well as the applied field (H) thus allowing us to study
systematically\cite{Mahi1,Moritomo,Kimura}. We will report a new form of\
insulator-metal transition in this series : The insulator-metal transition
is unsually unstable with respect to thermal cycling in absence and in
presence of a magnetic field.

\bigskip

\section{EXPERIMENT}

\bigskip

We measured four probe resistivity of \ polycrystalline Pr$_{0.5}$Ca$_{0.5}$%
Mn$_{1-x}$Cr$_{x}$O$_{3}$ (x = 0.015, 0.02, 0.025, and 0.03) in absence of
magnetic field and in presence of H = 2 T, and 5T. The temperature was
cycled down and up several times from the starting temperatures T$_{S}$ =
350 K, 300 K, 250 K, 160 K ,100 K, 60 K to the final temperature T$_{F}$ = 5
K at a constant sweep rate of 2.5 K/min. Magnetization under a weak magnetic
field of H = 1 mT\ and also under H = 2 T were measured down and up in
between T$_{S}$ = 300 K, 250 K,150 K and T$_{F}$ = 5 K.

\bigskip

\section{RESULTS AND\ DISCUSSIONS}

\bigskip

Figure 1(a) shows $\rho $(T) of x = 0.015 under H = 0 T when temperature is
cycled down and up from T$_{S}$ = 250 K to T$_{F}$ = 5 K. The number of
cycles is denoted by numeric 1,2,3 etc. The sample was not taken out the
cryostat prior to the completion of all the thermal cycles. After the 1$%
^{st} $ cycle 300 K$\rightarrow $5K$\rightarrow $300 K, T is reduced to T$%
_{S}$. In the 1$^{st}$ cycle, $\rho $ raises rapidly around 230 K due to
charge-orbital ordering and changes from insulating (d$\rho $/dT 
\mbox{$<$}%
0) to metal like(d$\rho $/dT 
\mbox{$>$}%
0) behavior below T$_{p}$ = 84 K. $\rho $(T) is hysteretic with warming
curve below and above the cooling curve for T $<$ T$_{p}$ and \ $>$ T$_{p}$
respectively and both curves merge for T 
\mbox{$>$}%
280 K. During the 2$^{nd}$ cycle, T$_{p}$ shifts down to 74 K and the value
of \ the resistivity $\rho _{p}$ at T$_{p}$ increases by more than a factor
of 10. Upon further thermal cycling, $\rho _{p}$ increases to $\approx $
1.95 x 10$^{5}$ and T$_{p}$ decreases to 45 K in the 4$^{th}$ cycle and $%
\rho $ exceeds more than 10$^{6}$ $\Omega $ cm at 40 K in the 5$^{th}$ and 6$%
^{th}$ cycles.\ This effect is dominant only for T 
\mbox{$<$}%
T$_{p}$ and T close but above T$_{p}$ but for T 
\mbox{$>$}%
100\ K, all $\rho $(T) curves overlap on each other. The variations of \ T$%
_{p}$ and $\rho _{p}$ with number of cycles will be presented latter in
Fig.3. The so obtained insulating state is stable and does not revert to the
original behavior (1$^{st}$ cycle) when the sample is heated to 300 K and
cycled between 300 K and 5 K several times. However, if the sample is air
annealed at 1273 K for five hours, \ the 1$^{st}$ cycle is reproducible with
identical T$_{p}$ and $\rho _{p}$. It should be mentioned that the virgin
sample was originally annealed at 1773 K \cite{Raveau} and annealing at 1273
K after thermal cycling is sufficient to remove any residual strain
developed. We found that a shorter annealing time, for example 3 hours at
1273 K, reintroduces the insulator metal transition but with different T$%
_{p} $ and $\rho _{p}$. A detailed annealing time dependence of \ the
insulator metal transition is under investigation and will be published in
future. For subsequent measurements to be rerported in this paper, we used
small bars (3mm x 1mm x 1mm) cut from a single big virgin pellet . Although
the insulator-metal transition in manganites is known to be sensitive to
chemical doping, pressure, magnetic field etc., $^{1-9}$ this is the first
time that a destruction of \ the insulator-metal transition just by thermal
cyling is reported in manganites. Our results mimics\ the behavior of (La$%
_{1-y}$A$_{y}$)$_{0.67}$Ca$_{0.33}$MnO$_{3}$ (y = Pr, Y etc) series whose T$%
_{p}$ decreases and $\rho _{p}$ increases with increasing \ content of
smaller Y rare earth ions\cite{Hwang}.

\bigskip

The observed effect is strongly dependent on the composition and decreases
with increasing x. Fig. 1(b) shows $\rho $ of x = 0.02. This compound
exhibits higher T$_{p}$ (= 120 K while cooling) and nearly ten times lower $%
\rho _{p}$ (= 4.8 $\Omega $ cm while cooling) than x = 0.015 (T$_{p}$ = 84 K
and $\rho _{p}$ = 136 $\Omega $ cm while cooling). There is a systematic
trend in the behavior of $\rho _{p}$ (warm) and T$_{p}$ (warm) with
increasing the number of cycles: $\rho _{p}$(warm) is comparable to $\rho
_{p}$(cool) in the 2$^{nd}$ cycles but becomes much lower than cooling \ as
the number of cycles increases. The weak shoulder in $\rho $(warm) which
develops around 105 K \ in the 3$^{rd}$ cycle becomes a prominent maximum in
the 10$^{th}$ cycle. As for x = 0.015, resistivity curves in different
cycles overlap on each other for T 
\mbox{$>$}%
\mbox{$>$}%
T$_{p}$. Compared to variations by several orders of magnitude in $\rho _{p}$
(cool) for x = 0.015, $\rho _{p}$ (cool) changes only by a factor of \ 5 in
x = 0.02,\ even in 10$^{th}$ cycle and by less than a factor of 0.8 in x =
0.03 (see Fig. 1(c)). The resistivity of x $\geq $ 0.03 is unaffected by
thermal cycling (not shown here). Fig. 1(d) shows $\rho $(T) of x = 0.015
under H = 2 T. We can clearly see the increase of the resistivity with
increasing the number of cycles and the insulator-metal transition is
abruptly lost in 8$^{th}$ cycle. A comparison \ of Fig. 1(a) and 1(d)
suggests that it is impossible to predict what should be the value of \
residual or peak resistivity before the sample loses metallic behavior. In
the rest of the paper we focus on x = 0.015 which shows the most dramatic
effect.

\bigskip

We also investigated \ whether the observed effect in x = 0.015 is dependent
on the starting temperature T$_{S}$. The resistivity curves for T$_{S}$ =
350 K, 300 K, and 200 K are identical to the data for T$_{S}$ = 250 K and
hence we do not present them here for the lack of space. The main panel of \
Fig. 2(a) shows $\rho $(T) for T$_{S}$ = 100 K and the inset shows data for T%
$_{S}$ = 160 K. The value of $\rho _{p}$ \ in the 5$^{th}$ cycle decreases
from $\approx $3 x 10$^{4}$ $\Omega $ cm for T$_{S}$ = 160 K to $\approx $%
575 $\Omega $ cm for T$_{S}$= 100 K and for T$_{S}$ = 60 K (not shown here)
whereas $\rho $ increases by less than a factor of 1.2. These results point
to the fact that the number of cycles needed to destroy insulator-metal
transition increases with lowering temperature below 200 K. Instead of
varying T$_{S}$, we also investigated the effect of varying T$_{F}$. Fig.
2(b) shows $\rho $(T) for x = 0.015 with T$_{S}$ = 250 K but T$_{F}$
decreasing from 100 K to 10 K instep of 10 K. In none of the cycles,
transition to metallic state is realized. Fig. 3 summarizes the variation of
T$_{p}$ and $\rho _{p}$ with the number of cycles and T$_{S}$ for x = 0.015
in presence and absence of magnetic field.

\bigskip

A doubt could be raised whether the increase of $\rho $ is caused by the
elapse of time during thermal cycling. To verify this point we measured time
dependence of the resistivity at 60 K\ in different cycles on a sample not
submitted to thermal cycling earlier. The result presented in the inset of
Fig. 2(a) in a normalized form ($\rho $(t)/$\rho $(t = 0), where $\rho $(t)
is the resistivity after t seconds) clearly suggests that $\rho $ decreases
with time as found by Kimura et al\cite{Kimura}. in Nd$_{0.5}$Ca$_{0.5}$Mn$%
_{0.98}$Cr$_{0.02}$O$_{3}$. This confirms that the observed increase of the
resistivity is purely a temperature driven effect. Thus, two different
antagonist mechanisms seems to act in thermal cycling and isothermal aging
effects. The isothermal aging effect is possibly caused by growth of
ferromagnetic clusters in size with time\cite{Kimura}.

\bigskip

To understand the origin of \ the unstable I -M transition induced by
thermal cycling, we carried out magnetization measurements. The main panel
of Fig. 4(a) shows M(T) of x = 0.015 under H = 1 mT for T$_{S}$ = 250 K. The
peak around 230 K is caused by charge-orbital ordering. Upon lowering
temperature further ferromagnetic clusters form below 150 K. Charges are
itinerant within these clusters by double exchange mechanism but the
resistivity does not decrease as these clusters are isolated. As T decreases
further intercluster interactions increase and \ when a percolating path
between ferromagnetic clusters is established around 84 K, $\rho $(T)
decreases as in Fig. 1(a). Fig. 4(a) clearly shows that the ferromagnetic
transition broadens in temperature and its phase fraction decreases each
time with temperature cycling. Interestingly, M decreases by about 30 \%
between the first two cycles \ but the overal change between the 1$^{st}$
and 5$^{th}$ cycles is only of about 46 \%. It should be noted that M does
not go to zero in the 5$^{th}$ cycle whereas metallic behavior is lost in
resistivity (see Fig. 1(a)). When T$_{S}$ = 160 K, the overall decrease
between the 1$^{st}$ and 5$^{th}$ cycles is only 19 \%, (see inset of Fig.
4(a)).

\bigskip

Fig. 4(b) shows M(T) measured under a field of 2 T \ in the field cooled
mode and the inset shows M(H) at 10 K made under zero field cooled mode.
M(H) in the 1$^{st}$ cycle (see inset in Fig. 4(b)) shows a two phase
behavior, one due to gradual alignment of ferromagnetic clusters in the
field direction below 1.5 T and another due to field induced canting of
spins in antiferromagnetic background above 1.5 T. The observed total
magnetization is M$_{obs}$ = f$_{m}$M$_{FM}$+(1-f$_{m}$)M$_{AF}$ where f$%
_{m} $ and 1-f$_{m}$ are the volume fractions of \ the ferromagnetic and the
antiferromagnetic phases with saturation magnetic moments M$_{FM}$ and M$%
_{AFM}$ respectivly. The spin only saturation moment of the Mn site cation
is 3.485 $\mu _{B}$ if spins of Cr$^{3+}$,Mn$^{3+}$, Mn$^{4+}$ ions are
parallel and 3.395 $\mu _{B}$ if the moment of Cr$^{3+}$ ion is reversed as
suggested earlier\cite{Mahi1}. Assuming that the magnetization of the
ferromagnetic phase completely saturates at 1.5 T, and the field induced
canting of antiferromagnetic moment is negligible below this field, we can
estimate the ferromagnetic phase fraction f$_{m}$ = M$_{obs}$/M$_{FM}$
=0.582/3.485 =16.7 \%.which is slightly above the percolation limit of p$%
_{c}\approx $14.5 \% in the continuum model\cite{Weber}. In the 8$^{th}$
cycle, f$_{m}$ decreases to 8.4 \%.( M$_{obs}$/M$_{FM}$ = 0.2938/3.485 = 8.4
\%). The values of f$_{m}$ are 17.2 \% in the 1$^{st}$ cycle and 8.6 \% if
we take M$_{FM}$ = 3.395$\mu _{B}$. As a consequence of the decrease of f$%
_{m}$, percolation is lost and so is the metallic character of the
resistivity at low temperatures. Including a finite value for M$_{AF}$ due
to canting at 1.5 T could modify the actual value of f$_{m}$, but the
decrease of the ferromagnetic phase fraction is still undisputable.

\bigskip

When the sample is field cooled under H = 2 T (main panel of Fig. 4(b)), the
ferromagnetic phase fraction is 19.6 \% and it reduces to 14.2 \% ( $\approx 
$ p$_{c}$) in the 8$^{th}$ cycle when the sample becomes insulator. These
results clearly suggest that the increase of resistivity is related to the
decrease of the ferromagnetic phase fraction. Since f$_{m}$ increases with
lowering T below $\thicksim $150 K in the 1$^{st}$cycle, the effect of
thermal cycling upon the increase of resistivity also decreases with T$_{S}$%
, lowering below 160 K.

\bigskip

At this moment a doubt could be raised : Is the decrease of M due to random
freezing of \ moments of ferromagnetic clusters by local anisotroy fields
instead of change in the ferromagnetic phase fraction ?. If cluster freezing
is the dominant mechanism, the anisotropy energy is expected to increase in
conjunction with the increase in resistivity and we can anticipate the
maximum in \ the zero field cooled magnetization (ZFC) to shift up in the
temperature with cycling\cite{Mydosh}. To verify this point we carried out
the ZFC -M(T) in H = 1 mT and H = 2 T. \ Fig. 5(a) shows \ M(T) under H = 1
mT in the ZFC mode after cooling the sample in zero field from 250K to 10 K
each time. The data were taken over a smaller temperature range compared to
Fig. 4(a) but it serves our purpose. In the first cycle, M(T) increases
smoothly from 10 K and exhibits a maximum around 57 K. However, in the
second cycle M(T) decreases dramatically in value the temperature range 20
K-110 K and in the next two cycles the change is rather small. The dramatic
decrease of M value in the 2$^{nd}$ cycle is not expected \ in a
conventional cluster or spin glass\cite{Mydosh} We do not observe the
anticipated upward shift of the maximum instead, the ZFC-M(T) curves in the
second, third, and fourth cycles show a maximum at T $\approx $ 30 K. Fig.
5(b) shows M(T) under H = 2 T in the ZFC mode. All the curves again exhibit
a maximum at the same temperature (T = 30 K) which also allows us to
conclude that cluster freezing is not the dominant mechanism in reducing the
value of the magnetization, but that the decrease in the ferromagnetic phase
fraction is the the possible origin.

\bigskip

What could be the mechanism which drives the diminution of the ferromagnetic
phase ? The ferromagnetic metallic (FMM) and the charge ordered insulating
(COI) phases are expected to differ structurally because of \ the larger
degree of Janh-Teller distortion in the latter phase and its absence in the
former phase. Then, inhomogeneous strains exist in the interfacial regions.
In \ a first order transition in which elastic interactions between the
parent and the product phases are non negligible (example martensitic
transition\cite{Cao}), growth of \ the product phase does not progress as a
function of time as one would expect, but needs changes in the thermal
energy. The rapid decrease in the magnetization between the first two cycles
and the gradual decrease in the later cycles resemble the evolution of
strain in the first few thermal cycles (known also as 'training') in
athermal martensitic materials\cite{Bigeon}. In charge ordered phase, d$%
_{Z^{2}}$ orbital ordering due to the Jahn-Teller distortion creates
anisotropic strain. We observed changes neither in intensity nor in the
value of the charge-orbital modulation vector {\bf q} by electron microscopy
at 92 K under thermal cycling (not shown here), however we cannot exclude
possible changes in the interfacial regions. We also carried out an X-ray
diffraction study at T = 300 K and at T = 100 K which is the lowest
temperature limit in our X-ray instrument (Philips- X'pert) on a piece of a
virgin sample and on a sample which has become an insulator after thermal
cycling. We carried out the Rietveld analysis in {\it pbmn} settings. The
results are presented in tables I and II. Although we find changes in the
structural parameters between the two samples, errors in the parameters (see
the numbers within brackets ) are too large to be reliable.\ It is possible
that structural changes occur mostly in the interfacial regions between the
ferromagnetic and the charge ordered phase which could not be detected in
the present X-ray study \ due to smaller width of the interfacial regions.
Millis et al\cite{Millis2} showed that biaxial strain due to the Jahn-Teller
distortion could lead to localization of charges and decrease\ the Curie
temperature. Then, a plausible explanation is that upon thermal cycling \
the Jahn-Teller distortion of \ Mn$^{3+}$O$_{6}$ octahedras in the
interfacial regions increases;as a result the interfacial elastic energy
increases which impedes the growth of the ferromagnetic phase during thermal
cycling similar to the martensitic transition\cite{Cao}. We do not think
that granularity effects play any important role in our results since we
find a systematics in the behavior of \ the thermal cycling effect with
different x although all the compounds were prepared under identical
conditions. Very recently, similar thermal cycling induced phenomena but
less dramatic effect than our results was also reported\ in Gd$_{5}$(Si$%
_{1.95}$Ge$_{0.25}$)\cite{Levin}.

\bigskip

\section{\protect\bigskip CONCLUSIONS\ }

In conclusion, we have shown for the first time that the insulator-metal
transition in the phase separated Pr$_{0.5}$Ca$_{0.5}$Mn$_{1-x}$Cr$_{x}$O$%
_{3}$ manganites is unstable with respect to thermal cycling. While the
resistivity much above the I-M temperature is found to be unaffected, the
resistivity close and below the I-M transition is strongly affected by
repeated thermal cycling. Repeated thermal cycling is found to increase the
low temperature resistivity and even destroys the metallic transition in x =
0.015. The effect is strongly dependent on the composition and decreases
with increasing x. We have shown that the ferromagnetic phase fraction
decreases with repeated thermal cycling. Our X-ray and electron diffraction
studies do not show any clear evidence for change in structural parameters
of the bulk, although changes in interfacial regions can not be excluded.We
suggest that increase ofthe elastic energy in the ferromagnetic-charge
ordered interface might be the origin of the instability of the I-M
transition. In contrary \ to the thermal cycling effect, resistivity at a
given temperature is found to decrease slowly as a function of time. Further
experimental and theoretical studies are necessary to clarify the origin of
the observed effects. It is also worth to investigate whether the observed
effect is restricted to only Cr-doped charge ordered manganite or is is the
generic behavior of the manganites for certain doping levels.

\bigskip

Acknowledgments:

R. M. thanks MENRT (France) for financial assistance and thanks Professor.
A. K. Raychaudhuri, Professor. M. R. Ibarra, and Professor. D. I. Khomskii
for useful discussions.

\begin{center}
\smallskip \newpage

FIGURE CAPTIONS
\end{center}

\begin{description}
\item[Fig. 1]  : Resistivity of Pr$_{0.5}$Ca$_{0.5}$Mn$_{1-x}$Cr$_x$O$_3$ in
number of thermal cycling under H = 0 T \ for x = 0.015 (a), 0.02 (b), 0.03
(c) and under H = 2 T for x = 0.015 (d). T$_S$ = starting temperature for
thermal cycling. Arrows indicate the direction of temperature sweep.

\item[Fig. 2]  : Resistivity of Pr$_{0.5}$Ca$_{0.5}$Mn$_{0.985}$Cr$_{0.015}$O%
$_3$ in different thermal cycles under H = 0 T for T$_S$ = 100 K (main
panel) and T$_S$= 160 K (inset). (b) Resistivity under partial thermal
cycling from a fixed T$_S$ = 250 K to a varying T$_F$ from 100 K by a step
of -10 K. Inset : Time dependence of resistivity at T = 60 K in different
thermal cycles.

\item[Fig. 3]  : Variation of T$_p$ (a) and $\rho _p$ (b) with number of
cyles during cooling (closed symbols) and warming (open symbols) for T$_S$ =
250 K under H = 0 T, 2 T and 5 T for x = 0.015. O(1) and O(2) respectively
stand for T$_S$ = 160 K and 100 K for H = 0 T.

\item[Fig. 4]  : Temperature dependence magnetization of Pr$_{0.5}$Ca$_{0.5}$%
Mn$_{0.985}$Cr$_{0.015}$O$_3$ under H = 1 mT for T$_S$= 250 K(a) and T$_S$ =
160 K.(inset). M(T) under 2 T in 1$^{st}$ and 8$^{th}$ cycle (b) and field
dependence of magnetization in 1$^{st}$ and 8$^{_{th}}$ cycles (inset). H$_c$
-critical field for metamagnetic transition.

\item[Fig. 5]  : M(T) of Pr$_{0.5}$Ca$_{0.5}$Mn$_{0.985}$Cr$_{0.015}$O$_3$
under zero field cooled mode for (a) H = 1 mT and (b) H = 2 T.

\newpage 
\end{description}

\begin{center}
TABLE CAPTIONS

\bigskip
\end{center}

\begin{description}
\item[Table 1]  Variation of lattice parameters and atomic positions of the
virgin and the cycled Pr$_{0.5}$Ca$_{0.5}$Mn$_{0.985}$Cr$_{0.015}$O$_3$
samples in {\it Pbmn} settings. The numbers in the bracket indicate error.

\item[Table 2]  : Variation of bond distances and bond angles of the virgin
and the cycled Pr$_{0.5}$Ca$_{0.5}$Mn$_{0.985}$Cr$_{0.015}$O$_3$ samples in 
{\it Pbmn} settings. The numbers in the bracket indicate error.
\end{description}

\end{document}